\documentclass{svjour3}
\usepackage{graphicx}

\begin{document}

\title{Enhancement of sudden death of entanglement for driven qubits
\thanks{This work was supported by the Academy of Finland (Acad. Res. Fellowship 00857 and
projects 7111994 and 7118122).
}}

\author{Jian Li, K.~Chalapat, and G.~S.~Paraoanu}

\institute{Low Temperature Laboratory, Helsinki University of Technology,\\
P.O. Box 5100, FIN-02015 TKK, Finland}

\date{\today}

\maketitle

\begin{abstract}

We study the recently discovered phenomena of sudden death of entanglement for a system of
two qubits, each of them independently longitudinally damped  by a reservoir and subjected to a continuous driving.
We show that
driving produces, in the interaction picture, an effective bath that has elements
amounting to various extra sources of noise (transverse, thermal squeezed, thermal longitudinal).
As a result, the time of sudden death decreases due to driving, which we verify as well by direct numerical calculation.
We suggest that this phenomenon can be studied systematically using superconducting qubits driven by microwave fields.



\end{abstract}

\section{Introduction}
\label{intro}

Decoherence is an ubiquitous phenomenon in quantum mechanics and any architecture of future quantum computers
must consider it seriously. Recently, it has been noted that already at the level of
two qubits, the effect of a weak noise on each qubit is not simply additive when it comes to nonlocal properties.
This effect, called "sudden death of entanglement" \cite{Yu1}, means that the concurrence of two qubits has an abrupt decay in time
rather than the exponential decay typical for single qubits. In this paper we consider the case of two qubits coupled independently
to two reservoirs, and driven coherently and resonantly by two continuous fields. We find that two-qubit states which in the absence of the drive would have
exponential decay now exhibit sudden death
of entanglement.

Although the results in this paper are valid for any type of qubits, we will briefly analyze the case of superconducting-qubit
architectures, where microwave fields
applied to each qubit can be used to manipulate the quantum two-qubit state, creating entanglement, CNOT and iSWAP gates, {\it etc.}
\cite{paraoanu}. With the emergence, during the last decade, of a plethora of such superconducting devices
as serious contenders for the role of the quantum bit, the effect of decoherence in these systems
should be thoroughly investigated. While for the case of single superconducting qubits this has been already
done to a large extend
both experimentally and theoretically,
there is yet no systematic experimental study of decoherence in superconducting many-qubit systems. Also,
circuit boards containing two superconducting qubits could be used as a workbench for fundamental
studies of decoherence effects.

%In other setups, such as those
%involving a resonator \cite{yale} or another qubit \cite{niskanen},
%the coupling is a residual one, in others \cite{fixed} it plays an essential role in realizing the quantum gates.

\section{The model}
\label{model}

We study a general model, similar to that of Ref. \cite{Yu1}, which can be adapted in a straigthforward way to any species of superconducting
qubits. We take $\hbar = 1$, and we consider two non-interacting qubits which are coupled to two uncorrelated thermal reservoirs:
\begin{eqnarray}
&& H_{sys} = -\sum_{j=1,2} \frac{\omega_j^L}{2}\sigma_j^z , \ \ \ \ H_{res} = \sum_{j=1,2} \sum_\lambda \omega_{j\lambda} a_{j\lambda}^\dag a_{j\lambda} , \nonumber \\
&& H_{s-r} = \sum_{j=1,2} \sum_\lambda \left( g_{j\lambda}^\ast a_{j\lambda}^\dag \sigma_j^- + g_{j\lambda} a_{j\lambda} \sigma_j^+ \right) , \label{eq_hamiltonians}
\end{eqnarray}
where $\omega_j^L$ denotes the Larmor frequency of the qubit
$j$, $\sigma_j^- = |\uparrow_j\rangle \langle\downarrow_j|$ ($\sigma_j^+ = |\downarrow_j\rangle \langle\uparrow_j|$)
is the qubit lowering (raising) operator, and $a_{j\lambda}^\dag$ ($a_{j\lambda}$) denotes the
 bosonic creation (annihilation) operator for the reservoir coupled to each qubit, respectively.
 Each qubit is irradiated with an external monochromatic microwave field. The qubit-field interaction is taken
  to be of the form
\begin{eqnarray}
H_{ext}(t) = \sum_{j = 1,2} \Omega_j(t)\cos(\omega_j t)\sigma_j^x , \label{eq_driving_terms}
\end{eqnarray}
with $\omega_j$ the angular frequency of the field applied to qubit-$j$, and $\Omega_j$ the
qubit-field coupling strength (Rabi frequency).

In the {\em weak external field approximation} \cite{Puri} $||H_{ext}|| \ll ||H_{sys}||$, which is valid in most
experiments with superconducting qubits, and zero temperature, the dynamical evolution of the qubits' density operator
$\rho_s$ is governed by the Born-Markov master equation
\begin{eqnarray}
\dot{\rho}_s = -i [H_{sys}, \rho_s] -i [H_{ext}, \rho_s] + {\cal L}[\rho_s] . \label{eq_original_master_equation}
\end{eqnarray}
The Liouvillean
\begin{eqnarray}
{\cal L}[\rho_s] = \sum_{j=1,2} \frac{\Gamma_j}{2} \left( 2\sigma_j^- \rho_s \sigma_j^+ - \sigma_j^+ \sigma_j^- \rho_s - \rho_s \sigma_j^+ \sigma_j^- \right)
\end{eqnarray}
models the longitudinal dampings of the qubits, and $\Gamma_j$ is the standard energy relaxation rate.

The dynamics of the two-qubit system given by Eq. (\ref{eq_original_master_equation}) is rather complex. However,
it is simplified considerably by transforming into the rotating reference frame, and performing a standard rotating wave
approximation (RWA) to eliminate the counter rotating terms. The master equation in the rotating frame becomes
\begin{equation}
\dot{\rho}_{rf} = -i [H_{rf}, \rho_{rf}] + {\cal L}[\rho_{rf}] , \label{eq_master_equation_in_rf}
\end{equation}
with $\rho_{rf} = S^\dag \rho_s S$, $S = \exp[i(\omega_1\sigma_1^z + \omega_2\sigma_2^z)t / 2]$, and
\begin{eqnarray}
H_{rf} &=& S^\dag (H_{sys} + H_{ext}) S + i\frac{\partial S^\dag}{\partial t}S \nonumber \\
&\approx& \sum_{j=1,2}\left( \frac{\delta_j}{2}\sigma_j^z + \frac{\Omega_j(t)}{2}\sigma_j^x \right) . \label{eq_h_rf}
\end{eqnarray}
Here $\delta_j = \omega_j - \omega_j^L$ is the detuning of the driving frequency from the corresponding qubit transition frequency.

\section{Discussion}
\label{discuss}

As in other studies on entanglement sudden death in two-qubit systems such as \cite{Yu2,Al},
we assume that the initial state of our system is an \lq\lq{\sf X} state\rq\rq
\begin{equation}
\rho_{rf}(0) = \rho_s(0) =
\left( \begin{array}{cccc}
        a_0 & 0 & 0 & w_0 \\
        0 & b_0 & z_0 & 0 \\
        0 & z_0^\ast & c_0 & 0 \\
        w_0^\ast & 0 & 0 & d_0
\end{array} \right) . \label{eq_initial_state}
\end{equation}
However, despite the fact that the \lq\lq{\sf X} state\rq\rq is a rather general form of a density matrix, due
to the existence of driving fields, the evolution under Eq. (\ref{eq_master_equation_in_rf}) does not preserve this form, resulting in
a rather complicated kinetic equation for the density matrix elements.
Thus, we need to find a way to reduce the number of coupled differential equations.

For the sake of simplicity, we assume on-resonance driving ($\delta_j = 0$) and time independent $\Omega_j$ (rectangular pulses),
then we move to the interaction picture by means of $\tilde{\rho} = \exp(iH_{rf}t)\rho_{rf}\exp(-iH_{rf}t)$.
The equation of motion for $\tilde{\rho}$ is thus
\begin{equation}
\dot{\tilde{\rho}} = \sum_{j=1,2} \frac{\Gamma_j}{2}\left( 2\tilde{\sigma}_j^- \tilde{\rho} \tilde{\sigma}_j^+ - \tilde{\sigma}_j^+ \tilde{\sigma}_j^- \tilde{\rho} - \tilde{\rho} \tilde{\sigma}_j^+ \tilde{\sigma}_j^- \right) , \label{eq_master_equation_in_ip}
\end{equation}
where

\begin{eqnarray}
\tilde{\sigma}_j^- &=& e^{i\Omega_j\sigma_j^x t/2} \sigma_j^- e^{-i\Omega_j\sigma_j^x t/2} =
-\frac{i}{2}\sin(\Omega_jt)\sigma_j^z + \sin^2\left(\frac{\Omega_jt}{2}\right)\sigma_j^+ +
\cos^2\left(\frac{\Omega_jt}{2}\right)\sigma_j^- , \label{eq_sigma_m_in_ip} \\
\tilde{\sigma}_j^+ &=& e^{i\Omega_j\sigma_j^x t/2} \sigma_j^+ e^{-i\Omega_j\sigma_j^x t/2} =
\frac{i}{2}\sin(\Omega_jt)\sigma_j^z + \cos^2\left(\frac{\Omega_jt}{2}\right)\sigma_j^+ +
\sin^2\left(\frac{\Omega_jt}{2}\right)\sigma_j^- . \label{eq_sigma_p_in_ip}
\end{eqnarray}

Substituting (\ref{eq_sigma_m_in_ip}) and (\ref{eq_sigma_p_in_ip}) into (\ref{eq_master_equation_in_ip}), the master equation in the interaction picture is explicitly given by (for notation simplicity, we drop the tilde over $\rho$ from now on)

\begin{eqnarray}
\dot{\rho} &=& \frac{i}{4}\sum_{j=1,2}\Gamma_j \sin(\Omega_jt) \left( \sigma_j^+ \rho\sigma_j^z + \sigma_j^- \rho\sigma_j^z - \sigma_j^z \rho \sigma_j^+ - \sigma_j^z \rho\sigma_j^- + \sigma_j^+\rho + \rho\sigma_j^+ - \sigma_j^-\rho - \rho\sigma_j^- \right) \nonumber \\
&& + \frac{i}{8}\sum_{j=1,2}\Gamma_j \sin(2\Omega_jt) \left( \sigma_j^z \rho\sigma_j^- + \sigma_j^- \rho\sigma_j^z - \sigma_j^z \rho \sigma_j^+ - \sigma_j^+ \rho\sigma_j^z \right) \nonumber \\
&& + \frac{1}{16}\sum_{j=1,2} \Gamma_j [ 3 + 4\cos(\Omega_jt) + \cos(2\Omega_jt) ] \left( 2\sigma_j^-\rho\sigma_j^+ - \sigma_j^+\sigma_j^- \rho - \rho\sigma_j^+\sigma_j^- \right) \nonumber \\
&& + \frac{1}{16}\sum_{j=1,2} \Gamma_j [ 3 - 4\cos(\Omega_jt) + \cos(2\Omega_jt) ] \left( 2\sigma_j^+\rho\sigma_j^- - \sigma_j^-\sigma_j^+ \rho - \rho\sigma_j^-\sigma_j^+ \right) \nonumber \\
&& + \frac{1}{8}\sum_{j=1,2} \Gamma_j [ 1-\cos(2\Omega_jt) ] \left( \sigma_j^+ \rho\sigma_j^+ + \sigma_j^-\rho\sigma_j^- + \sigma_j^z \rho\sigma_j^z - \rho \right) . \label{eq_explicit_master_equation_in_ip}
\end{eqnarray}

It is easy to verify that when the external fields are very weak, $\Omega_j\ll \Gamma_j$, even in a relatively long time scale $\tau\propto \max[\Gamma_j^{-1}]$, Eq. (\ref{eq_explicit_master_equation_in_ip}) is reduced to
\begin{eqnarray}
\dot{\rho}\approx \sum_{j=1,2} \frac{\Gamma_j}{2}\left( 2\sigma_j^- \rho\sigma_j^+ - \sigma_j^+\sigma_j^-\rho - \rho\sigma_j^+\sigma_j^- \right) . \label{small}
\end{eqnarray}

A more realistic situation is the so-called secular
limit $\Omega_j\gg \Gamma_j$ \cite{Cohen}. For instance, in experiments with quantronium circuits \cite{chargeflux},
the Rabi frequency $\Omega_j$ is usually larger than 100 MHz, whereas the relaxation rate $\Gamma_j < 2$ MHz.
In this case, we may neglect all the oscillating terms with frequencies $\Omega_j$ and $2\Omega_j$ in
Eq. (\ref{eq_explicit_master_equation_in_ip}), and arrive at a simple equation of motion

\begin{eqnarray}
&&\dot{\rho} \approx  \sum_{j=1,2} \frac{\Gamma_j}{8} \left( \sigma_j^+\rho\sigma_j^+ + \sigma_j^-\rho\sigma_j^- + \sigma_j^z\rho\sigma_j^z - \rho \right) \nonumber \\
&&+ \sum_{j=1,2}\frac{3\Gamma_j}{16}\left( 2\sigma_j^-\rho\sigma_j^+ + 2\sigma_j^+\rho\sigma_j^- - \sigma_j^+\sigma_j^-\rho - \sigma_j^-\sigma_j^+\rho - \rho\sigma_j^+\sigma_j^-
- \rho\sigma_j^-\sigma_j^+ \right) . \label{eq_final_master_equation}
\end{eqnarray}

We note that Eq. (\ref{eq_final_master_equation}) can be regarded as describing losses
caused by a combination of three uncorrelated effective noise sources of each qubit:

\paragraph{transverse thermal reservoir}
\begin{eqnarray}
{\cal L}_{aj}[\rho] = \frac{\Gamma_j}{8}\left( \sigma_j^z\rho\sigma_j^z - \rho \right) ;
\end{eqnarray}

\paragraph{longitudinal high temperature thermal reservoir}
\begin{eqnarray}
{\cal L}_{bj}[\rho] &=& \frac{\gamma_j}{2} (\bar{n}_j + 1) \left( 2\sigma_j^-\rho\sigma_j^+ - \sigma_j^+\sigma_j^-\rho - \rho\sigma_j^+\sigma_j^- \right) \nonumber \\
&& + \frac{\gamma_j}{2}\bar{n}_j \left( 2\sigma_j^+\rho\sigma_j^- - \sigma_j^-\sigma_j^+\rho - \rho\sigma_j^-\sigma_j^+ \right) ,
\end{eqnarray}
with $\bar{n}_j + 1 \approx \bar{n}_j$ and $\gamma_j \bar{n}_j / 2 \approx \Gamma_j / 8$;

\paragraph{longitudinal squeezed vacuum reservoir}\cite{Scully}
\begin{eqnarray}
{\cal L}_{cj}[\rho] &=& \frac{\kappa_j}{2} \left[ \cosh^2(r_j) \left( 2\sigma_j^-\rho\sigma_j^+ - \sigma_j^+\sigma_j^-\rho - \rho\sigma_j^+\sigma_j^- \right) \right. \nonumber \\
&& \ \ \ \ \ \left. + \sinh^2(r_j) \left( 2\sigma_j^+\rho\sigma_j^- - \sigma_j^-\sigma_j^+\rho - \rho\sigma_j^-\sigma_j^+ \right) \right] \nonumber \\
&& - \kappa_j\sinh(r_j)\cosh(r_j) \nonumber \\
&& \ \ \ \ \ \times \left( e^{-i\theta_j}\sigma_j^- \rho \sigma_j^- + e^{i\theta_j}\sigma_j^+\rho\sigma_j^+ \right) ,
\end{eqnarray}
with $\theta_j = \pm \pi$ being the reference phase for the effective squeezed field, $r_j$ being the
squeeze parameter which fulfills $\sinh(r_j)\approx\cosh(r_j)$ (large squeezing limit) and $\kappa_j\sinh(r_j)\cosh(r_j)\approx \Gamma_j/8$.

In other words,
\begin{equation}
\dot{\rho}=\sum_{j=1,2}{\cal L}_{aj}[\rho]+{\cal L}_{bj}[\rho]+{\cal L}_{cj}[\rho]
\end{equation}

Therefore even without further calculations, we can predict that the external driving fields will enhance the sudden death
effect, i.e., it will destroy the long-lived entangled states investigated by Yu and Eberly \cite{Yu1},
due to the raising of reservoir's effective temperature \cite{Al}, and combination of noises \cite{Yu3}.

\section{Sudden death time}
\label{eds_time}

Because of its invariance under local unitary transforms, the concurrence of the system can be estimated by solving
Eq. (\ref{eq_final_master_equation}), which is much simpler than solving (\ref{eq_master_equation_in_rf}) since the {\sf X} states remains the {\sf X} form in the evolution given by (\ref{eq_final_master_equation}). Since $\rho(0) = \rho_s(0)$, we may set
\begin{equation}
\rho(t) =
\left( \begin{array}{cccc}
        a(t) & 0 & 0 & w(t) \\
        0 & b(t) & z(t) & 0 \\
        0 & z^\ast(t) & c(t) & 0 \\
        w^\ast(t) & 0 & 0 & d(t)
\end{array} \right) , \label{eq_x_state}
\end{equation}
and by substituting it into (\ref{eq_final_master_equation}) we obtain the following kinetic equation for those non-zero density matrix elements:
\begin{eqnarray}
&& \dot{a} = 3[ \Gamma_1(c - a) + \Gamma_2(b - a) ]/8 , \nonumber \\
&& \dot{b} = 3[ \Gamma_1(d - b) + \Gamma_2(a - b) ]/8 , \nonumber \\
&& \dot{c} = 3[ \Gamma_1(a - c) + \Gamma_2(d - c) ]/8 , \nonumber \\
&& \dot{d} = 3[ \Gamma_1(b - d) + \Gamma_2(c - d) ]/8 , \nonumber \\
&& \dot{z} = [ \Gamma_1(w^\ast - 5z) + \Gamma_2(w - 5z) ]/8 , \nonumber \\
&& \dot{w} = [ \Gamma_1(z^\ast - 5w) + \Gamma_2(z - 5w) ]/8 . \label{eq_kinetic_equation}
% && \dot{w}^\ast(t) = \frac{1}{8}\left\{ \Gamma_1[z(t) - 5w^\ast(t)] + \Gamma_2[z^\ast(t) - 5w^\ast(t)] \right\} . \nonumber \\
\end{eqnarray}

By defining $\eta_j = \exp(-\Gamma_jt/4)$, the solution of Eq. (\ref{eq_kinetic_equation}) has the form:

\begin{eqnarray}
&& a = \left[a_0\left( 1 + \eta_1^3 + \eta_2^3 + \eta_1^3\eta_2^3 \right) + b_0\left( 1 + \eta_1^3 - \eta_2^3 - \eta_1^3\eta_2^3 \right) + c_0\left( 1 - \eta_1^3 + \eta_2^3 - \eta_1^3\eta_2^3 \right) + d_0\left( 1 - \eta_1^3 - \eta_2^3 + \eta_1^3\eta_2^3 \right) \right] / 4 , \nonumber \\
&& b = \left[a_0\left( 1 + \eta_1^3 - \eta_2^3 - \eta_1^3\eta_2^3 \right) + b_0\left( 1 + \eta_1^3 + \eta_2^3 + \eta_1^3\eta_2^3 \right) + c_0\left( 1 - \eta_1^3 - \eta_2^3 + \eta_1^3\eta_2^3 \right) + d_0\left( 1 - \eta_1^3 + \eta_2^3 - \eta_1^3\eta_2^3 \right) \right] / 4 , \nonumber \\
&& c = \left[a_0\left( 1 - \eta_1^3 + \eta_2^3 - \eta_1^3\eta_2^3 \right) + b_0\left( 1 - \eta_1^3 - \eta_2^3 + \eta_1^3\eta_2^3 \right) + c_0\left( 1 + \eta_1^3 + \eta_2^3 + \eta_1^3\eta_2^3 \right) + d_0\left( 1 + \eta_1^3 - \eta_2^3 - \eta_1^3\eta_2^3 \right) \right] / 4 , \nonumber \\
&& d = \left[a_0\left( 1 - \eta_1^3 - \eta_2^3 + \eta_1^3\eta_2^3 \right) + b_0\left( 1 - \eta_1^3 + \eta_2^3 - \eta_1^3\eta_2^3 \right) + c_0\left( 1 + \eta_1^3 - \eta_2^3 - \eta_1^3\eta_2^3 \right) + d_0\left( 1 + \eta_1^3 + \eta_2^3 + \eta_1^3\eta_2^3 \right) \right] / 4 , \nonumber \\
&& z = \eta_1^2\eta_2^2\left[ w_0\left( 1 + \eta_1 - \eta_2 - \eta_1\eta_2 \right) + z_0\left( 1 + \eta_1 + \eta_2 + \eta_1\eta_2 \right) + w_0^\ast\left( 1 - \eta_1 + \eta_2 - \eta_1\eta_2 \right) + z_0^\ast\left( 1 - \eta_1 - \eta_2 + \eta_1\eta_2 \right) \right] / 4 , \nonumber \\
&& w = \eta_1^2\eta_2^2\left[ w_0\left( 1 + \eta_1 + \eta_2 + \eta_1\eta_2 \right) + z_0\left( 1 + \eta_1 - \eta_2 - \eta_1\eta_2 \right) + w_0^\ast\left( 1 - \eta_1 - \eta_2 + \eta_1\eta_2 \right) + z_0^\ast\left( 1 - \eta_1 + \eta_2 - \eta_1\eta_2 \right) \right] / 4 . \nonumber \\ \label{eq_solution_of_kinetic_equation}
\end{eqnarray}

\begin{figure}[htb]
\includegraphics[width=9cm]{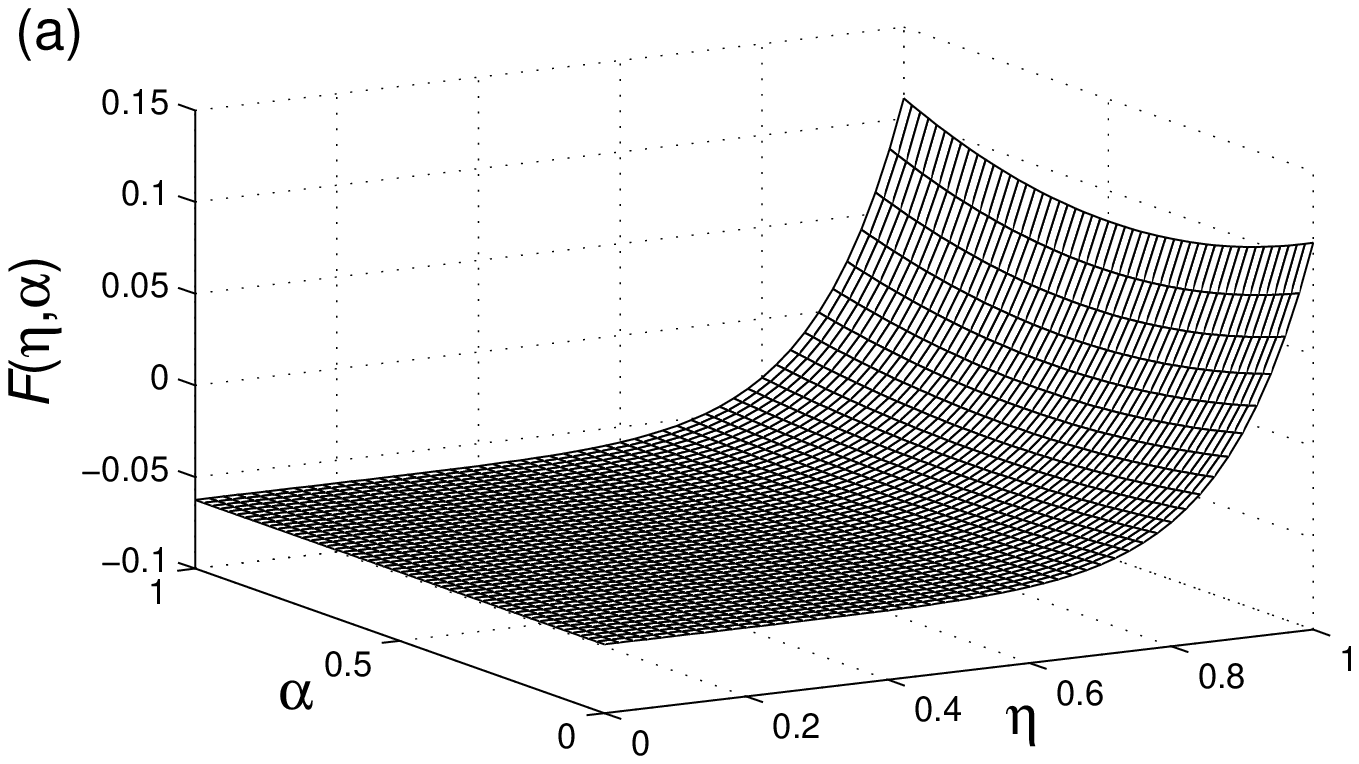}\includegraphics[width=9cm]{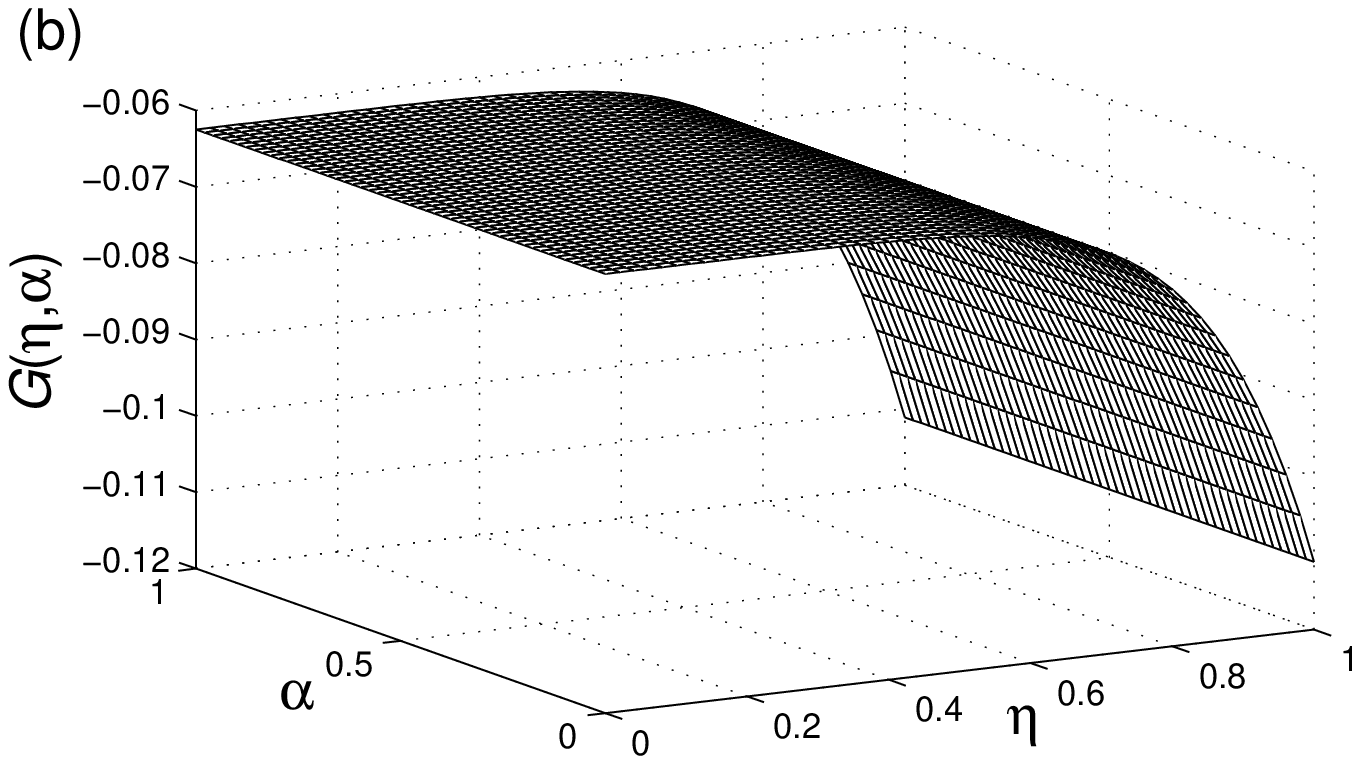}
\caption{$F$ and $G$ as functions of $\eta$ and $\alpha$. } \label{fig_F_x}
\end{figure}

To obtain the sudden death time, we follow Sec. III in \cite{Al}. For even greater simplicity, we assume $\Gamma_1 = \Gamma_2 \equiv \Gamma$, thus $\eta_1 = \eta_2 \equiv \eta$, and $0\leq \eta \leq 1$. The two equations for the sudden death time,
\begin{eqnarray}
&& F(t)\equiv |z(t)|^2 - a(t)d(t) = 0 \nonumber \\
\mathrm{and} && G(t)\equiv |w(t)|^2 - b(t)c(t) = 0 ,
\end{eqnarray}
are polynomial equations of degree 12 of $\eta$, which can not be solved straightforward.
However, as illustrated in Fig. \ref{fig_F_x}, by plotting $F$ and $G$ vs $\eta$ for an initial Yu-Eberly state
\begin{equation}
\rho_{YE} = \frac{1}{3}
\left( \begin{array}{cccc}
        1-\alpha & 0 & 0 & 0 \\
        0 & 1 & 1 & 0 \\
        0 & 1 & 1 & 0 \\
        0 & 0 & 0 & \alpha
\end{array} \right) , \label{eq_alpha_state}
\end{equation}
we find that both functions are continuous in the range $0\leq\eta\leq 1$, $G$ is always negative in this range, and $F$ becomes zero at $\eta\approx 0.9$ for all values of $\alpha$, which indicates the sudden death time $t_{sd}\approx 0.42/\Gamma$ for all $\alpha$. As we discussed in the preceding section, the external microwave fields enhance the disentanglement of long-lived entangled states with $\alpha\leq 1/3$.

\section{Numerical verification}
\label{numerical}

\begin{figure}[htb]
\includegraphics[width=9cm]{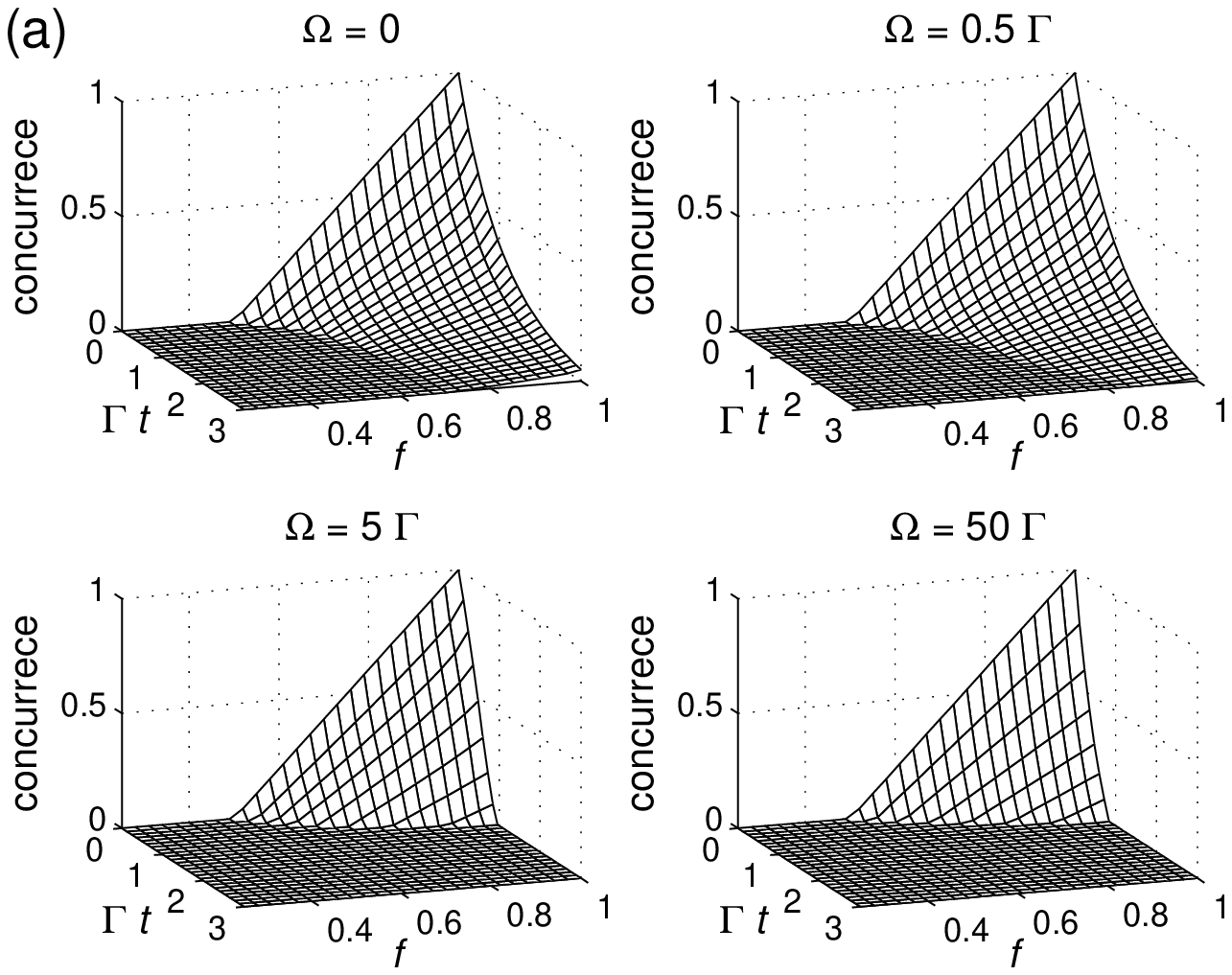}
\includegraphics[width=9cm]{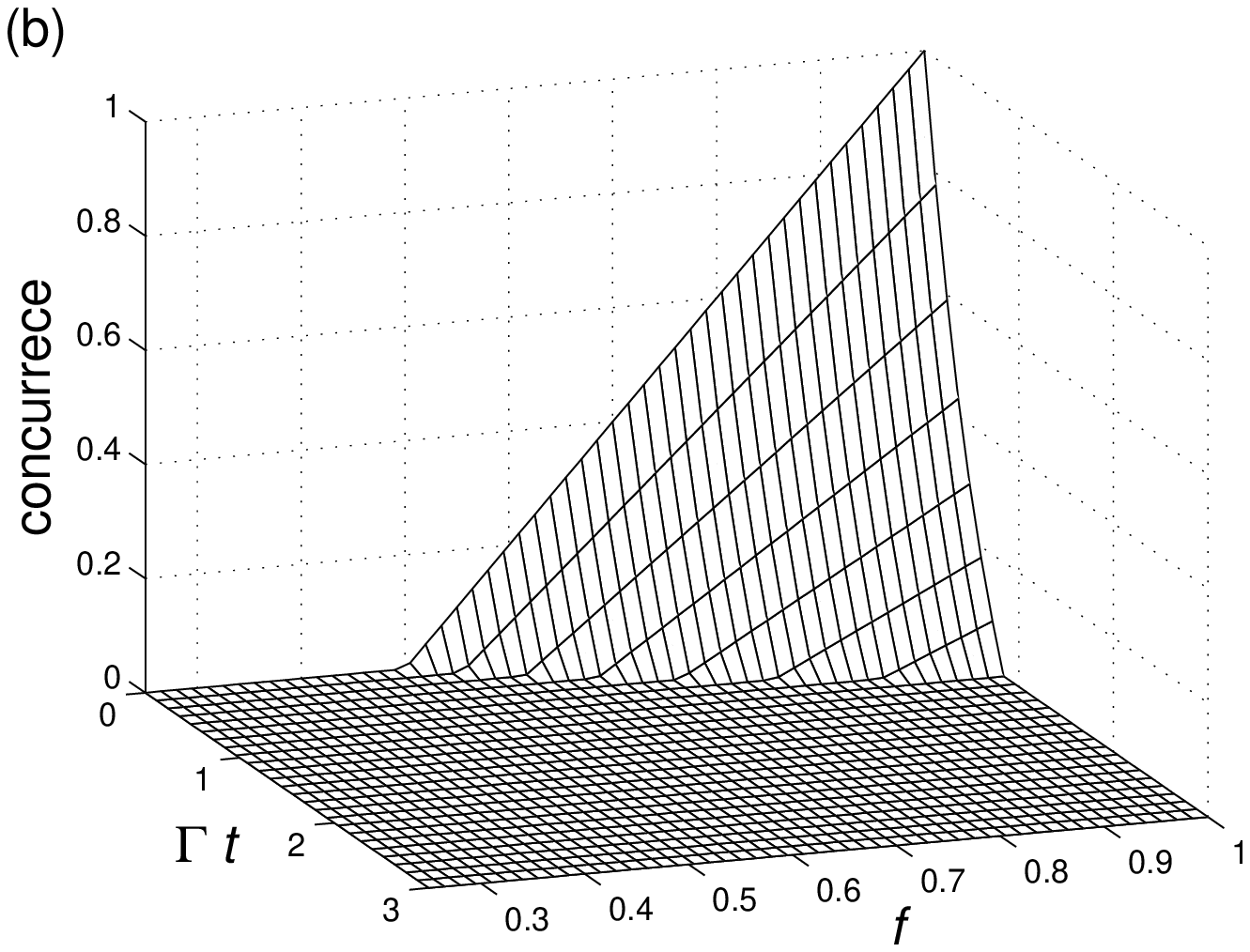}
\caption{(a) Time evolution of concurrence for the Werner state in the rotating frame (with four different driving amplitudes)
obtained by solving numerically the initial master equation Eq.(\ref{eq_master_equation_in_rf}).
One can notice that at low Rabi frequencies ($\Omega= 0$, $\Omega = 0.5\Gamma$) the sudden death effect appears only for
some of the states, as expected from \cite{Yu1} and Eq. (\ref{small}). However, in the regime of larger Rabi frequencies
($\Omega= 5\Gamma$, $\Omega = 50\Gamma$), the sudden death effect is enhanced.
(b) Time evolution of concurrence for the Werner state in the interaction picture and the limit of large Rabi frequencies ($\Omega = 50\Gamma$)
, obtained by using the analytical solution of the master equation Eq. (\ref{eq_final_master_equation}), namely
Eq.
(\ref{eq_solution_of_kinetic_equation})
and Eq. (\ref{eq_concurrence_formula}).
} \label{fig_concurrence}
\end{figure}
In Sec. \ref{discuss}, to derive the analytically solvable master equation (\ref{eq_final_master_equation}), we have made the
 approximation of neglecting all the oscillating terms. Now, we numerically check the validity of this approximation.
We consider that initially the two qubits are in a Werner state \cite{Yu2}
\begin{equation}
\rho_W = \frac{1}{3}
\left( \begin{array}{cccc}
        1-f & 0 & 0 & 0 \\
        0 & (1+2f)/2 & (1-4f)/2 & 0 \\
        0 & (1-4f)/2 & (1+2f)/2 & 0 \\
        0 & 0 & 0 & 1-f
\end{array} \right) , \label{eq_Werner_state}
\end{equation}
with the fidelity $0.25\leq f\leq 1$.

We first calculate the concurrence in the rotating frame with the original formula \cite{Wootters}
\begin{equation}
{\cal C}(\rho_{rf}) = \max\{ 0,\lambda_1-\lambda_2-\lambda_3-\lambda_4 \} ,
\end{equation}
where $\lambda_i$s are the eigenvalues of $\sqrt{\sqrt{\rho_{rf}} \widetilde{\rho}_{rf}\sqrt{\rho_{rf}}}$
in decreasing order, with $\widetilde{\rho}_{rf} \equiv (\sigma^y \otimes \sigma^y) \rho_{rf}^\ast (\sigma^y \otimes \sigma^y)$.
Fig. \ref{fig_concurrence}(a) shows the time evolution of ${\cal C}(\rho_{rf})$ obtained by numerically solving the master
equation (\ref{eq_master_equation_in_rf}). For simplicity, we take the Rabi frequencies $\Omega_1 = \Omega_2 \equiv \Omega$, and
still keep $\Gamma_1 = \Gamma_2 = \Gamma$. In the interaction picture, $\rho$ remains in the {\sf X} form during its time evolution,
 therefore a simplified formula \cite{Yu2,Al}
\begin{equation}
{\cal C}(\rho) = 2\max\{ 0,|z(t)|-\sqrt{a(t)d(t)}, |w(t)|-\sqrt{b(t)c(t)} \}
\label{eq_concurrence_formula}
\end{equation}
is applied for computing the concurrence. Fig. \ref{fig_concurrence}(b) illustrates the time evolution of
concurrence for the Werner state obtained by direct substituting (\ref{eq_solution_of_kinetic_equation})
into (\ref{eq_concurrence_formula}).

By comparing Fig. \ref{fig_concurrence}(b) with the lower subplots of Fig. \ref{fig_concurrence}(a), one can find that the approximation of neglecting oscillating terms is very good for large $\Omega$. From the upper subplots of Fig. \ref{fig_concurrence}(a) one can also find that the enhancement of sudden death can happen even when $\Omega < \Gamma$.

\section{Conclusion}
\label{conclusion}

In this paper we have studied the entanglement of two non-interacting or interacting qubits driven by external
on-resonance microwave fields, in the presence of uncorrelated zero temperature longitudinal noise sources.
We have found that in the secular limit, the behavior of the concurrence is very similar to that at finite temperature and
under both longitudinal and transverse dampings. Although the case of two qubits coupled to transverse noises was not
considered in this paper, they can be easily investigated in the same way as for longitudinal noises.
In the secular limit and in the interaction picture, the transverse noises will contribute with the following Liouvillean
\begin{eqnarray}
{\cal L}_\varphi[\rho] &\approx& \sum_{j=1,2}\frac{\Gamma_{\varphi j}}{4} \left( \sigma_j^z\rho\sigma_j^z - 2\rho + \sigma_j^-\rho\sigma_j^+ + \sigma_j^+\rho\sigma_j^- \right. \nonumber \\
&& \ \ \ \ \ \ \ \ \left. - \sigma_j^+\rho\sigma_j^+ - \sigma_j^-\rho\sigma_j^- \right).
\end{eqnarray}

\end{document}